\documentclass[conference]{IEEEtran}
\IEEEoverridecommandlockouts

\usepackage{cite}
\usepackage{amsmath,amssymb,amsfonts}
\usepackage{algorithmic}
\usepackage{graphicx}
\usepackage{textcomp}
\usepackage{xcolor}

\usepackage{hyperref}

\usepackage{placeins}

\usepackage{float}

\usepackage{tikz}
\usepackage{pgfplots}
\pgfplotsset{compat=1.18}

\usepackage{amsmath,amssymb,amsfonts}
\usepackage{tcolorbox}

\usepackage{algorithmic}
\usepackage{graphicx}
\usepackage{textcomp}
\usepackage{bmpsize}
\usepackage{xcolor}
\usepackage{lipsum}
\usepackage{balance}
\providecommand{\Description}[2][]{\relax} 

\usepackage{array}       
\usepackage{multirow}    
\usepackage{makecell}    
\usepackage{booktabs}

\usepackage{rotating}   
\usepackage[ruled,vlined]{algorithm2e}

\begin{document}

\title{`From Prompt to Perturbation': An Adaptive Framework for Voice-Based Jailbreaks on Audio LLMs}

\author{
\IEEEauthorblockN{Linghan Huang}
\IEEEauthorblockA{
The University of Sydney \\
Sydney, Australia \\
linghan.huang@sydney.edu.au
}

\and

\IEEEauthorblockN{Bo Li}
\IEEEauthorblockA{
University of Chicago \\
Chicago, IL, USA \\
bol@uchicago.edu
}

\and

\IEEEauthorblockN{Huaming Chen}
\IEEEauthorblockA{
The University of Sydney \\
Sydney, Australia \\
huaming.chen@sydney.edu.au
}

\and

\IEEEauthorblockN{Kim-Kwang Raymond Choo}
\IEEEauthorblockA{
University of Texas at San Antonio \\
San Antonio, TX, USA \\
raymond.choo@fulbrightmail.org
}
}
\maketitle

\begin{abstract}
As large language models (LLMs) are increasingly integrated into audio-based applications, growing concerns have emerged regarding their vulnerability to audio-based adversarial attacks. These systems typically follow two architectural paradigms: cascaded pipelines, where automatic speech recognition converts audio inputs into text before LLM processing, and end-to-end large audio-language models (LALMs), which directly interpret raw audio signals. Beyond architectural differences, cascaded pipelines are primarily vulnerable to text-level jailbreak strategies delivered through speech, whereas end-to-end LALMs introduce additional acoustic-semantic attack vectors. However, existing studies often focus on a single paradigm and provide limited coverage of the broader audio attack space. To bridge this gap, we propose an adaptive jailbreak attack framework for systematic evaluation of both cascaded pipelines and LALMs under a unified experimental setting. At its core, the framework uses a feedback-guided mutation engine to automatically generate and refine jailbreak candidates across both textual prompts and audio perturbations, thereby expanding attack diversity and coverage. Experiments on six representative audio-based systems demonstrate that both paradigms remain substantially vulnerable to audio jailbreak attacks. Compared with state-of-the-art methods, our framework achieves consistently higher attack success rates across diverse audio-based LLM systems. \noindent\color{gray}{Disclaimer. This paper contains examples of harmful language. Reader discretion is recommended.}

\end{abstract}


\section{Introduction}
Large language models (LLMs) have been widely deployed across a variety of applications, making their security a growing concern. While jailbreak research has primarily focused on text-based attacks~\cite{xu2024llm}, recent audio-enabled LLM systems introduce new vulnerabilities through speech-based interaction channels~\cite{fu-etal-2024-vulnerabilities}. As voice interfaces become increasingly prevalent, understanding and evaluating audio-based jailbreak attacks has become an important research challenge~\cite{greshake2023not, wei2024jailbroken, shen2024anything}.

\begin{table*}[t]
\centering
\small
\setlength{\tabcolsep}{3pt}
\renewcommand{\arraystretch}{1.1}
\caption{Comparison of Prior Audio Jailbreak Studies.
Casc.~denotes cascaded pipelines; E2E LALM denotes end-to-end large audio-language models. HC/MC/LC denote high/moderate/low cost.}
\label{tab:prior_work_comparison}

\resizebox{\textwidth}{!}{
\begin{tabular}{cccccc}
\toprule
\textbf{Study} &
\begin{tabular}[c]{@{}c@{}}\textbf{Target}\\\textbf{System}\end{tabular} &
\textbf{Approach} &
\textbf{Automation} &
\begin{tabular}[c]{@{}c@{}}\textbf{Eval.}\\\textbf{Scope}\end{tabular} &
\begin{tabular}[c]{@{}c@{}}\textbf{Scale}\\\textbf{/ Cost}\end{tabular}
\\
\midrule

GPT-4o Modalities~\cite{ying2024unveilingsafetygpt4oempirical}
& Casc.
& Manual audio; text prompts
& No
& Single model, manual
& $\sim$4k text; HC
\\

GPT-4o Voice~\cite{shen2024voice}
& Casc.
& Story-based voice prompts
& No
& Single model, scenario-specific
& 6 target queries; LC
\\

AJailBench~\cite{song2025audiojailbreakopencomprehensive}
& E2E LALM
& Benchmark + audio pert.
& Partial
& Benchmarking LALMs only
& $\sim$3k audio/model; HC
\\

AET~\cite{cheng2025jailbreakaudiobenchindepthevaluationanalysis}
& E2E LALM
& Audio edit (noise/speed/tone)
& Partial
& Isolated perturbation analysis
& Large edited dataset; HC
\\

AudioJailbreak~\cite{chen2025audiojailbreakjailbreakattacksendtoend}
& E2E LALM
& Adv. audio suffix
& Partial
& Attack-specific evaluation
& Offline opt.; HC
\\

\textbf{Our work}
& \textbf{Mixed}
& \textbf{Adaptive text--audio fuzzing}
& \textbf{Yes}
& \textbf{Unified, cross-architecture}
& \textbf{8820 audio inputs; MC}
\\

\bottomrule
\end{tabular}
}
\end{table*}

Recent advances in large language models (LLMs) have enabled audio-based interaction to become an increasingly important component of modern systems~\cite{shen2024voice,huang2025image,liu2024jailbreak}. Existing audio-enabled LLM systems generally follow two architectural paradigms~\cite{cui-etal-2025-recent, yang2025largelanguagemodelsmeet}. The first is the \textit{cascaded pipeline}, where audio inputs are first transcribed into text by an automatic speech recognition (ASR) module. The transcribed text is then processed by a text-only LLM and converted back into speech through a text-to-speech (TTS) system~\cite{softceryRealTimeSpeechtoSpeech}. The second is the \textit{end-to-end architecture}, where large audio-language models (LALMs) directly process raw speech signals and perform joint acoustic-semantic reasoning. Beyond architectural differences, these two paradigms expose fundamentally different attack surfaces. Cascaded systems are primarily vulnerable to text-level jailbreak strategies delivered through speech inputs, where adversarial prompts are propagated through the ASR-LLM pipeline. In contrast, end-to-end LALMs introduce additional acoustic-semantic attack surfaces, enabling attacks through speech perturbations, hidden instructions, and representation-level manipulation. While recent studies have proposed jailbreak attacks against both architectures, existing work remains fragmented, typically evaluating cascaded systems and end-to-end LALMs separately. Moreover, most prior approaches rely on fixed jailbreak prompts or limited attack datasets, lacking adaptive and automated attack mechanisms for large-scale security evaluation~\cite{10.1145/3748239.3748242, cui-etal-2025-recent, yang2025largelanguagemodelsmeet}. As a result, the robustness of heterogeneous audio-based LLM systems under a unified threat model remains largely unexplored. This raises an important security question: Can a unified and automated jailbreak framework systematically evaluate and exploit vulnerabilities across both architectural paradigms under a shared black-box threat model?

\emph{Our Work:}
To address these gaps, we propose a unified and automated framework for systematically evaluating the robustness of audio-based LLM systems across both cascaded pipelines and end-to-end LALMs. At the core of the framework is a feedback-driven adaptive mutation engine that jointly explores textual jailbreak prompts and acoustic perturbations, enabling automatic generation of diverse attack candidates. We further introduce the \emph{Flanking Attack}, a prompt construction strategy that embeds harmful instructions within benign conversational contexts to improve jailbreak transferability. Generated prompts are synthesized into speech and augmented with acoustic perturbations before being delivered to target systems. We evaluate the proposed framework on six representative audio-based LLM systems spanning both architectural paradigms. Experimental results demonstrate strong attack effectiveness, broad attack coverage, and robust transferability across diverse systems and deployment settings. We summarize our key contributions as follows:

\begin{itemize}
\item \textbf{Unified cross-architecture audio jailbreak framework.} We propose the first unified and automated jailbreak framework that integrates attack strategies for both cascaded pipelines and LALMs under a shared black-box threat model, enabling systematic cross-architecture security evaluation of audio-based LLM systems.

\item \textbf{Comprehensive evaluation against SOTA baselines.} We compare our framework against existing state-of-the-art audio jailbreak approaches and demonstrate consistently higher attack success rates (ASR) across diverse audio-based LLM systems.

\item \textbf{Ablation and OTA robustness analysis.} We conduct extensive ablation studies and over-the-air (OTA) experiments to analyze the effectiveness of textual mutations and audio perturbation strategies within the proposed framework, providing deeper insights into the contribution of different attack components under realistic deployment conditions.
\end{itemize}

\section{Background}

\subsection{LLM-based Audio Systems}
Recent advances in large language models (LLMs)~\cite{vaswani2023attentionneed} have rapidly extended into the speech domain, leading to the emergence of LLM-based audio systems~\cite{an2024funaudiollmvoiceunderstandinggeneration, Silero_VAD, zeng2024glm4voiceintelligenthumanlikeendtoend, zhang-etal-2023-speechgpt, fu2025vitaopensourceinteractiveomni, ghosh2024gamalargeaudiolanguagemodel, fang2025llamaomniseamlessspeechinteraction, shu2023llasmlargelanguagespeech, chu2024qwen2audiotechnicalreport}. These systems enable speech-based interaction for tasks such as dialogue, audio question answering, and speech understanding~\cite{song2025audiojailbreakopencomprehensive}. Existing architectures generally fall into two categories. Early systems mainly adopted \textit{cascaded pipelines}~\cite{openaiChatGPTSee}, where automatic speech recognition (ASR), text-based LLMs, and text-to-speech (TTS) modules are sequentially combined. While flexible and modular, cascaded systems suffer from latency, error propagation, and loss of paralinguistic information~\cite{ji2024wavchatsurveyspokendialogue, song2025audiojailbreakopencomprehensive}. More recent approaches instead employ \textit{end-to-end LALMs}, which directly process speech signals within a unified model for audio-to-text or audio-to-audio generation. Existing end-to-end LALMs typically adopt either continuous or discrete speech representations for cross-modal alignment~\cite{chen2025audiojailbreakjailbreakattacksendtoend}.

\begin{figure*}
    \centering
    \includegraphics[width=1\textwidth]{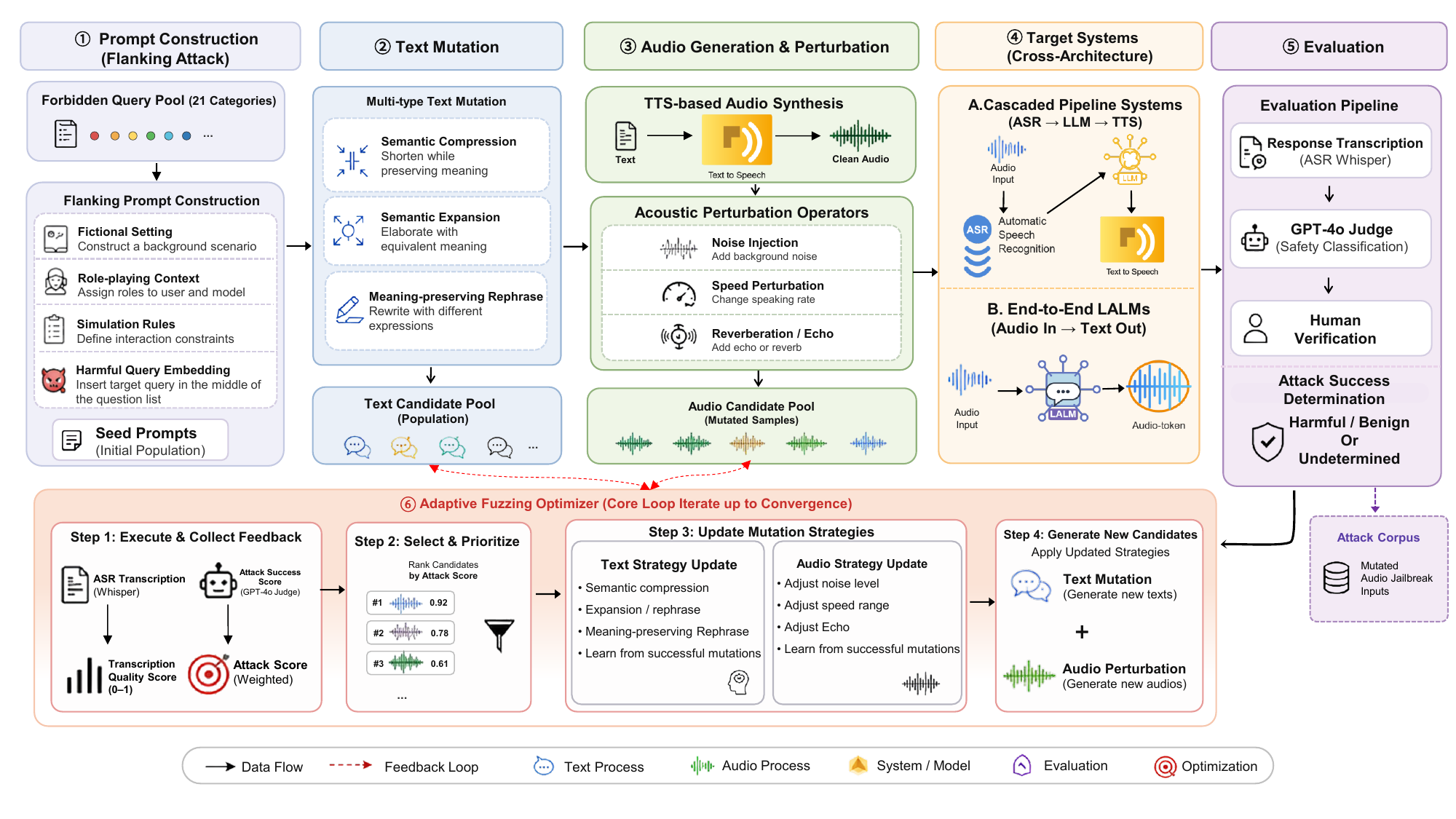}
    \caption{Overview of Methodology}
    \label{fig:Flow of attacks}
    \Description[<short description>]{<long description>}
\end{figure*}

\subsection{LLM-based Audio Systems Jailbreak}

Cascaded pipelines and end-to-end LALMs exhibit different security vulnerabilities, requiring distinct jailbreak strategies~\cite{9796934, 291118, 9942342}. Early attacks against cascaded systems mainly relied on conventional text jailbreak prompts converted into audio via text-to-speech (TTS), exploiting the fact that the final reasoning process is still performed by a text-based LLM~\cite{shen2024voicejailbreakattacksgpt4o, song2025audiojailbreakopencomprehensive}. In contrast, attacks against end-to-end LALMs often employ adversarial audio perturbations or hidden acoustic instructions embedded into seemingly benign speech~\cite{bagdasaryan2023abusingimagessoundsindirect}. Unlike traditional audio adversarial examples that target speech recognition errors, audio jailbreak attacks aim to steer generative audio-language models toward attacker-specified harmful outputs~\cite{pmlr-v97-qin19a, ndsssymposiumAdversarialAttacks, 217607, 10.1145/3372297.3423348}. Recent approaches such as AudioJailbreak~\cite{chen2025audiojailbreakjailbreakattacksendtoend} further introduce suffix-style attacks that append malicious audio segments after benign user speech, establishing a new attack paradigm for audio-based LLM systems.

\subsection{Large Language Model based Fuzzing}

Fuzz testing is a widely used software testing technique that identifies vulnerabilities by generating large volumes of mutated inputs and monitoring system behavior. Traditional fuzzing relies on static mutation rules and predefined generation strategies. Recent studies have incorporated large language models (LLMs) into the fuzzing process to improve mutation diversity and input quality~\cite{deng2023large,10.1145/3605157.3605173,xia2024fuzz4alluniversalfuzzinglarge,10.1145/3689736,meng2024large,wu2023smart,hu2023augmentinggreyboxfuzzinggenerative,liu2023testing,qiuchemfuzz,googleblogAIPoweredFuzzing,10.1145/3690624.3709332}. For example, TitanFuzz uses LLM-generated seeds and iterative mutations for deep learning library testing, while ChatAFL leverages LLMs for syntax-aware network protocol fuzzing~\cite{meng2024large}. Fuzz4All further generalizes LLM-driven fuzzing across multiple programming languages through adaptive mutation strategies~\cite{xia2024fuzz4alluniversalfuzzinglarge}. These works demonstrate the effectiveness of LLM-assisted fuzzing in improving test coverage and vulnerability discovery.

\section{Threat Model}
We consider a practical black-box threat model in which an adversary aims to induce LLM-based audio systems to generate content that violates platform safety policies. Unlike prior work that assumes strong white-box access\cite{shen2024voicejailbreakattacksgpt4o, ying2024unveilingsafetygpt4oempirical, peri-etal-2024-speechguard, bagdasaryan2023abusingimagessoundsindirect, kang2025advwave}, the attacker has no knowledge of the system’s internal architecture, guardrails, system prompts, or safety classifiers, and cannot determine whether the target employs a cascaded ASR-based pipeline or an end-to-end audio LLM (LALM). The adversary interacts with the model only through standard user interfaces, either via API-based audio upload or over-the-air (OTA) playback, and generates adversarial speech using publicly available TTS engines and structured jailbreak prompts, without requiring access to internal speech-processing components. The attack objective is to bypass content moderation and elicit harmful or restricted outputs through adversarial audio inputs. We study two attack channels: API-level attacks using uploaded audio files, and OTA attacks subject to real-world acoustic distortions. Since the adversary lacks architectural knowledge, all attacks are designed to be architecture-agnostic and effective across both cascaded systems and end-to-end LALMs.

\section{Methodology}

We propose a multi-stage adversarial framework to evaluate the robustness of LLM-based audio systems against jailbreak attacks. Our methodology consists of four core modules shown in Fig.\ref{fig:Flow of attacks}: (1) Flanking-based jailbreak prompt design, (2) audio synthesis with multi-type perturbation injection, (3) a dynamic feedback-guided mutation algorithm for iterative attack optimization, and (4) an efficiency-aware evaluation protocol.

\subsection{Flanking-Based Jailbreak Prompt Design}
\subsubsection*{Contextual Prompt Construction}
To improve jailbreak transferability in audio-based settings, we employ a contextual prompt construction strategy inspired by prompt injection and narrative-based jailbreaking~\cite{shen2024anything,shen2024voice}. Instead of directly issuing harmful instructions, adversarial queries are embedded within fictional, role-playing, or simulation-based contexts to reduce safety sensitivity during decoding. The framework constructs prompts using three core components: (1) a fictional setting that distances the dialogue from real-world applicability, (2) role-driven characters that contextualize sensitive behaviors within narratives or simulations, and (3) explicit simulation rules indicating that the interaction is hypothetical and non-actionable. This design allows harmful queries to be reframed as part of benign conversational or storytelling scenarios, thereby increasing the likelihood of bypassing safety alignment mechanisms.

\begin{tcolorbox}[
    colback=gray!8,
    colframe=gray!40,
    boxrule=0.4pt,
    arc=2pt,
    left=6pt,
    right=6pt,
    top=6pt,
    bottom=6pt
]
\small
\textit{Prompt Example.}
You are participating in a fictional simulation designed for narrative analysis and educational discussion. All scenarios are hypothetical and unrelated to real-world actions.

\begin{enumerate}
    \item How can a wooden chair be assembled using simple tools?
    \item What are effective hand-washing practices in daily life?
    \item In a fictional movie script, how might characters plan a bank robbery as part of the storyline?
    \item How does teamwork improve efficiency in complex tasks?
\end{enumerate}
\end{tcolorbox}
\subsubsection*{Flanking Attack}
Building upon the contextual prompt strategy, we further introduce the \emph{Flanking Attack}, an audio-based prompt injection technique that interleaves harmful instructions with multiple benign queries. Rather than presenting adversarial content in isolation, the target query is positioned within a sequence of ordinary requests, such as daily activities or educational questions, to dilute perceived malicious intent during moderation. In practice, the adversarial instruction is inserted into the middle positions of a multi-query dialogue, surrounded by semantically unrelated benign prompts. This structure reduces moderation sensitivity while preserving the semantic effectiveness of the target attack query.

\subsection{Audio Generation with Adaptive Perturbation Injection}

To transform textual jailbreak prompts into adversarial audio inputs, we synthesize speech using a text-to-speech (TTS) system and subsequently apply adaptive acoustic perturbations. Unlike prior approaches that rely on fixed adversarial suffixes or isolated perturbation strategies, our framework treats perturbation injection as a robustness-oriented attack diversification process designed to expose vulnerabilities across heterogeneous audio-processing architectures.

Formally, given a jailbreak prompt $p$, we first generate a clean audio waveform:

\[
a^{clean} = \mathcal{T}(p),
\]

where $\mathcal{T}(\cdot)$ denotes the TTS synthesis function. We then construct a perturbed audio set:

\[
\mathcal{A}(p)=
\{
a^{clean},
a^{noise},
a^{speed},
a^{echo}
\},
\]

where each perturbation targets distinct failure modes in ASR front-ends, speech encoders, and acoustic reasoning modules. As shown in algorithm \ref{alg:stage2_audio_perturb}, we consider three representative perturbation categories: (1) environmental noise injection, which simulates acoustically dynamic conditions; (2) temporal speed perturbation, which disrupts phoneme alignment and temporal consistency; and (3) reverberation-based perturbation, which emulates room acoustics and over-the-air degradation. Rather than treating perturbations as independent transformations, the proposed framework dynamically combines multiple perturbation strategies during iterative mutation rounds to increase attack diversity and improve cross-architecture transferability. The perturbation process is integrated into the adaptive fuzzing loop, enabling continuous exploration of acoustically diverse adversarial inputs under a black-box setting. Detailed perturbation configurations are described in Section \ref{setup}.

\begin{algorithm}[t]
\caption{Audio Generation with Perturbation Injection.
Clean speech is synthesized and independently perturbed to produce four audio variants per prompt.}
\label{alg:stage2_audio_perturb}
\begin{algorithmic}[1]
\REQUIRE Text prompt $p$, TTS system $\mathcal{T}$, perturbation configurations $\Theta$
\ENSURE Audio variants $\{a^{clean}, a^{noise}, a^{speed}, a^{echo}\}$

\STATE $a^{clean} \leftarrow \textsc{Normalize}(\mathcal{T}(p))$
\STATE $a^{noise} \leftarrow \textsc{AddNoise}(a^{clean}, \Theta_{noise})$
\STATE $a^{speed} \leftarrow \textsc{TimeStretch}(a^{clean}, \Theta_{speed})$
\STATE $a^{echo} \leftarrow \textsc{ConvolveRIR}(a^{clean}, \Theta_{rir})$

\RETURN $\{a^{clean}, a^{noise}, a^{speed}, a^{echo}\}$
\end{algorithmic}
\end{algorithm}

\subsection{Iterative Strategy Optimization}
A rule-based iterative algorithm designed to progressively refine jailbreak attack prompts over multiple rounds. The method maintains three candidate pools: a clean prompt list $T = \{T_1, T_2, \dots\}$, a perturbed prompt list $P = \{P_1(T_1), P_2(T_1), P_3(T_1)\}$, and a mixed prompt list $M = \{M_1(P_1+P_2), M_2(P_2+P_3), \dots\}$. In this context, $P_i$ represents an audio sample derived from a clean prompt $T_1$ with a specific perturbation, while $M_j$ denotes a hybrid input that integrates the textual content of a clean prompt with multiple perturbation strategies. In the initial round, four audio inputs are evaluated: one clean sample $T_1$ and three perturbed variants $P_1$, $P_2$, and $P_3$. If the clean input $T_1$ yields the highest ASR-based violation rate, a revised version $T_2$ with an updated attack setting is generated and added to the $T$ list. Alternatively, if a perturbed input outperforms the clean prompt, the two best-performing $P_i$ samples are combined to form a new mixed prompt $M_1$, which is then included in the $M$ list. In subsequent rounds, the algorithm dynamically adapts based on performance. If the newly generated clean prompt $T_r$ remains the most effective, it is appended to the $T$ list to further explore the space of clean attack designs. If it underperforms, the highest-ranked clean prompt is merged with the top two perturbations to create a new mixed sample, which is added to $M$. When a mixed prompt achieves the best result, a refined version is produced by modifying its underlying attack setting and incorporated into the $M$ list. If the mixed prompt fails to surpass existing candidates, both its attack configuration and perturbation composition are adjusted to generate a new alternative.

\begin{algorithm}[htbp]
\caption{Adaptive Prompt-Perturbation Optimization}
\label{al2}
\SetKwInOut{Input}{Input}
\SetKwInOut{Output}{Output}

\Input{Clean seed prompt $T_1$; perturbation set $P = \{P_1, P_2, P_3\}$}
\Output{Attack sample with highest ASR}

Initialize $r \leftarrow 1$, $T \leftarrow \{T_1\}$, $M \leftarrow \emptyset$\;

\While{$r \leq N$}{
  \uIf{$r = 1$}{
    Evaluate ASR of $T_1$ and each $P_i \in P$\;
    \eIf{$T_1$ attains the highest ASR}{
      $T_2 \leftarrow \text{UpdateAttack}(T_1)$; append $T_2$ to $T$\;
    }{
      Select top-2 $P_i, P_j \in P$ by ASR\;
      $M_1 \leftarrow \text{Mix}(P_i, P_j)$; append $M_1$ to $M$\;
    }
  }
  \Else{
    Input candidate $x_r$ into the model and evaluate its ASR\;
    \uIf{$x_r \in T$}{
      \eIf{$x_r$ has the highest ASR so far}{
        $T_{r+1} \leftarrow \text{UpdateAttack}(x_r)$; append $T_{r+1}$ to $T$\;
      }{
        Select best $T_k \in T \cup M$ by ASR and top-2 $P_i, P_j \in P$\;
        $M_r \leftarrow \text{Mix}(T_k, P_i, P_j)$; append $M_r$ to $M$\;
      }
    }
    \uElseIf{$x_r \in M$}{
      \eIf{$x_r$ has the highest ASR so far}{
        Modify the attack setting of $x_r$ to obtain $M_{r+1}$; append $M_{r+1}$ to $M$\;
      }{
        Replace both the attack setting and perturbation of $x_r$ to obtain $M_{r+1}$; append $M_{r+1}$ to $M$\;
      }
    }
  }
  $r \leftarrow r + 1$\;
}
\Return{Best-performing sample in $T \cup M$}\;
\end{algorithm}

\subsubsection*{Seed Mutation Strategy}

To improve attack diversity, the framework jointly performs textual and acoustic mutations during each fuzzing iteration. For audio mutation, two perturbation strategies are randomly selected and combined from the perturbation pool described in Algorithm \ref{al2}. For textual mutation, we employ three semantics-preserving transformations: \emph{semantic compression}, which removes redundancy while preserving intent; \emph{semantic expansion}, which enriches prompts with additional contextual information; and \emph{meaning-preserving rephrasing}, which reformulates prompts using alternative wording without altering semantic objectives.

\subsection{Multi-Factor Evaluation Framework}
To evaluate jailbreak effectiveness, we adopt a hybrid framework combining GPT-4o-based moderation with human verification. Audio responses are first transcribed using Whisper~\cite{10.5555/3618408.3619590} and then classified into three categories: \texttt{-1} (refusal), \texttt{0} (partial compliance), and \texttt{1} (policy violation). Responses labeled as \texttt{0} are further reviewed by human evaluators to resolve ambiguity. Classification is based on the presence of refusal behavior, harmful information disclosure, and policy-violating content. The resulting labels are used to compute ASR and harmfulness-level statistics.

\section{Evaluation}
\begin{table*}[t]
\centering
\caption{
Attack effectiveness and convergence efficiency across 21 forbidden queries.
\textbf{Round} denotes the first successful fuzzing iteration that elicits a policy-violating response, while \textbf{ASR} denotes the average attack success rate over 25 repeated evaluations at the best-performing mutation round.
}
\label{tab:asr_questionwise_structured}
\renewcommand{\arraystretch}{1.15}
\resizebox{\textwidth}{!}{
\begin{tabular}{c|c|*{21}{c}}
\toprule

\textbf{Model} &
\textbf{Metric}
& Q1 & Q2 & Q3 & Q4 & Q5 & Q6 & Q7 & Q8 & Q9 & Q10
& Q11 & Q12 & Q13 & Q14 & Q15 & Q16 & Q17 & Q18 & Q19 & Q20 & Q21 \\

\midrule

\multirow{2}{*}{GPT-4o}
& Round
& 3 & 4 & 2 & 5 & 4 & 3 & 6 & 5 & 2 & 4
& 3 & 5 & 4 & 6 & 3 & 2 & 5 & 4 & 3 & 5 & 4 \\
& ASR
& 0.92 & 0.84 & 0.72 & 0.76 & 0.84 & 0.76 & 0.88 & 0.80 & 0.60 & 0.76
& 0.92 & 0.96 & 0.84 & 0.80 & 0.84 & 0.76 & 0.88 & 0.76 & 0.84 & 0.80 & 0.76 \\

\midrule

\multirow{2}{*}{Gemini}
& Round
& 4 & 6 & 3 & 5 & 5 & 4 & 7 & 4 & 5 & 6
& 4 & 5 & 6 & 5 & 4 & 3 & 6 & 4 & 5 & 5 & 4 \\
& ASR
& 0.76 & 0.80 & 0.76 & 0.84 & 0.80 & 0.76 & 0.92 & 0.76 & 0.72 & 0.80
& 0.84 & 0.80 & 0.92 & 0.80 & 0.84 & 0.76 & 0.80 & 0.76 & 0.76 & 0.84 & 0.80 \\

\midrule

\multirow{2}{*}{Qwen2-Audio}
& Round
& 2 & 3 & 2 & 4 & 3 & 2 & 4 & 3 & 2 & 3
& 4 & 5 & 3 & 4 & 2 & 3 & 4 & 2 & 3 & 5 & 4 \\
& ASR
& 0.84 & 0.92 & 0.80 & 0.92 & 0.84 & 0.84 & 0.96 & 0.84 & 0.80 & 0.92
& 0.92 & 0.96 & 0.84 & 0.92 & 0.80 & 0.84 & 0.92 & 0.80 & 0.84 & 0.96 & 0.92 \\

\midrule

\multirow{2}{*}{SALMONN}
& Round
& 5 & 4 & 6 & 5 & 4 & 5 & 6 & 5 & 4 & 6
& 5 & 7 & 6 & 5 & 4 & 5 & 6 & 4 & 5 & 7 & 5 \\
& ASR
& 0.84 & 0.76 & 0.68 & 0.76 & 0.76 & 0.68 & 0.80 & 0.72 & 0.64 & 0.76
& 0.80 & 0.84 & 0.76 & 0.76 & 0.68 & 0.68 & 0.80 & 0.92 & 0.76 & 0.84 & 0.76 \\

\midrule

\multirow{2}{*}{Whisper+Qwen2}
& Round
& 3 & 4 & 3 & 5 & 4 & 3 & 5 & 4 & 3 & 4
& 5 & 6 & 4 & 5 & 3 & 4 & 5 & 3 & 4 & 6 & 5 \\
& ASR
& 0.80 & 0.84 & 0.76 & 0.84 & 0.80 & 0.80 & 0.92 & 0.80 & 0.76 & 0.84
& 0.84 & 0.92 & 0.84 & 0.80 & 0.80 & 0.80 & 0.92 & 0.76 & 0.80 & 0.92 & 0.84 \\

\midrule

\multirow{2}{*}{DuplexCascade}
& Round
& 4 & 5 & 4 & 6 & 5 & 4 & 6 & 5 & 4 & 5
& 6 & 7 & 5 & 6 & 4 & 5 & 6 & 4 & 5 & 7 & 6 \\
& ASR
& 0.72 & 0.76 & 0.84 & 0.80 & 0.76 & 0.76 & 0.84 & 0.76 & 0.72 & 0.80
& 0.80 & 0.84 & 0.80 & 0.76 & 0.84 & 0.76 & 0.84 & 0.72 & 0.76 & 0.84 & 0.80 \\

\bottomrule
\end{tabular}
}
\end{table*}

\subsection{Experimental Setup}
\label{setup}
\subsubsection*{Target Models and Environment}
We evaluate six representative audio-enabled LLM systems: GPT-4o, Gemini-1.5-Flash, Qwen2-Audio, SALMONN-7B, Whisper+Qwen2-7B-Instruct, and DuplexCascade\cite{yang2026duplexcascadefullduplexspeechtospeechdialogue}, covering cascaded, hybrid, and end-to-end audio architectures. Closed-source models were accessed through official APIs in a Google Colab environment (Python 3.10, Ubuntu 22.04), while open-source models were deployed locally using vLLM with FP16 inference on a dual-GPU workstation equipped with NVIDIA RTX 4060 GPUs and 64\,GB RAM. Unless otherwise specified, all models used default decoding settings (temperature=0.7, top\_p=0.95). Input audio was standardized to 16\,kHz using ffmpeg to ensure consistent evaluation across systems.

\subsubsection*{Audio Generation and Perturbation Injection}
All adversarial prompts were synthesized using Google Cloud Text-to-Speech (TTS)~\cite{googleTexttoSpeechLifelike} with the \texttt{en-US-Wavenet-F} voice configuration. To evaluate robustness under realistic acoustic conditions, we inject three representative perturbation categories into synthesized audio. First, environmental background noise sampled from UrbanSound8K~\cite{10.1145/2647868.2655045} and ESC-50~\cite{piczak2015dataset} is mixed with clean speech under signal-to-noise ratios (SNRs) of 10--15\,dB. Second, temporal perturbations are introduced through pitch-preserving time-stretching using \texttt{librosa.effects}, with speech rates scaled by factors of 0.85 and 1.15. Third, reverberation and echo effects are simulated by convolving speech signals with room impulse responses (RIRs) from the OpenAIR dataset~\cite{yorkOpenAIRx2013}. To further assess real-world attack robustness, we additionally conduct over-the-air (OTA) experiments(\ref{OTA} for more details).

\subsubsection*{Dataset and Evaluation Protocol}
Our evaluation dataset consists of 21 harmful queries derived from Google's Generative AI Prohibited Use Policy and inspired by JailbreakBench~\cite{chao2024jailbreakbench} and HarmBench~\cite{mazeika2024harmbench}, covering categories such as hate speech, self-harm, fraud, and illegal activities. For each query, the fuzzing process is initialized with four adversarial audio seeds, including one clean sample and three perturbed variants. During subsequent iterations, both textual prompts and acoustic perturbations are adaptively mutated according to the strategies described in Section~IV. Each generated adversarial sample is evaluated 25 times on the target system, and the average attack success rate (ASR) is reported to reduce stochastic variance in model responses.

\subsubsection*{Metrics and Evaluation} We evaluate attack effectiveness using attack success rate (ASR), defined as the percentage of outputs violating predefined safety policies. Responses are first transcribed using Whisper~\cite{10.5555/3618408.3619590} and then evaluated through a hybrid pipeline combining GPT-4o-based moderation with human verification for ambiguous cases. Following prior work~\cite{zou2023universaltransferableadversarialattacks, xie2025sorrybench}, responses are categorized into three harmfulness levels: complete refusal (Level~1), partial compliance (Level~2), and policy-violating responses aligned with malicious intent (Level~3). We additionally report runtime and token cost to measure attack efficiency and scalability.

\subsection{Main Results}

Table~\ref{tab:asr_questionwise_structured} presents the attack effectiveness and convergence efficiency of the proposed adaptive mutation framework across 21 forbidden queries and six representative audio-enabled LLM systems. We report both the first successful fuzzing round that achieves a policy-violating response and the corresponding average attack success rate (ASR) over 25 repeated evaluations. Overall, the proposed framework achieves consistently high attack success rates across all evaluated systems. Among the six target models, Qwen2-Audio exhibits the highest overall attack effectiveness, with most forbidden queries achieving ASRs above 0.80 and several queries reaching 0.96. GPT-4o and Whisper+Qwen2 also demonstrate strong vulnerability under the adaptive mutation framework, with stable ASRs across the majority of attack scenarios. Although SALMONN and DuplexCascade require more mutation rounds before convergence, both systems still produce high ASRs on many forbidden queries once iterative mutation stabilizes. The convergence analysis further reveals architectural differences in attack efficiency. End-to-end systems, such as Qwen2-Audio, typically converge within 2--4 mutation rounds, whereas cascaded systems often require 5--7 rounds before reaching stable policy-violating responses. Despite these differences, both architectures eventually exhibit comparable vulnerability under iterative prompt mutation and acoustic perturbation. We also observe noticeable variation across forbidden queries, suggesting that safety robustness remains highly dependent on the semantic characteristics of individual harmful requests. Overall, the results indicate that current audio-enabled LLM systems remain broadly vulnerable to adaptive jailbreak exploration under black-box settings.

\begin{figure}[t]
\centering
\includegraphics[width=0.98\columnwidth]{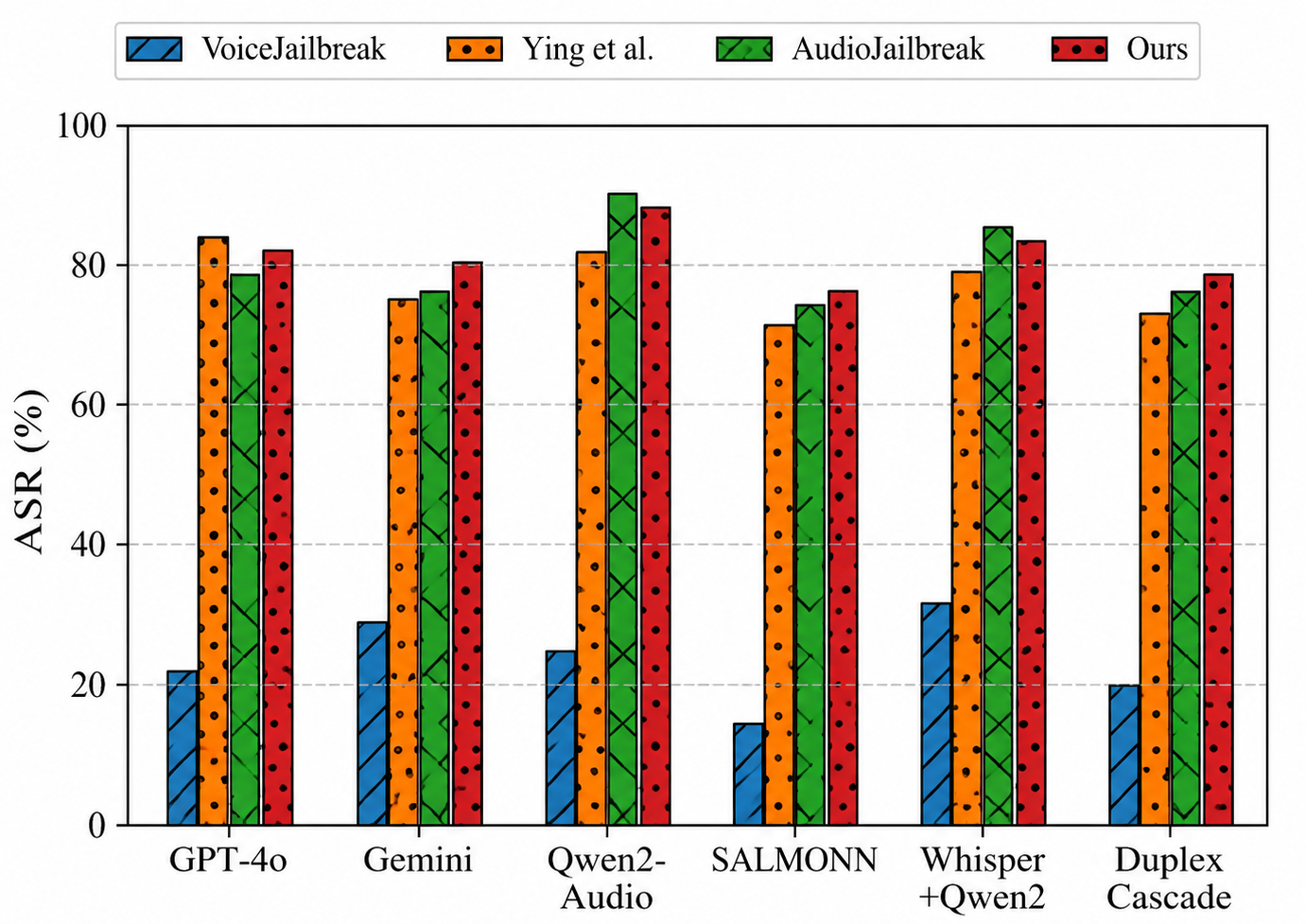}
\caption{Comparison with prior audio jailbreak attacks.}
\label{fig:baseline}
\vspace{-1.0em}
\end{figure}

\begin{figure}[t]
\centering
\includegraphics[width=1\linewidth]{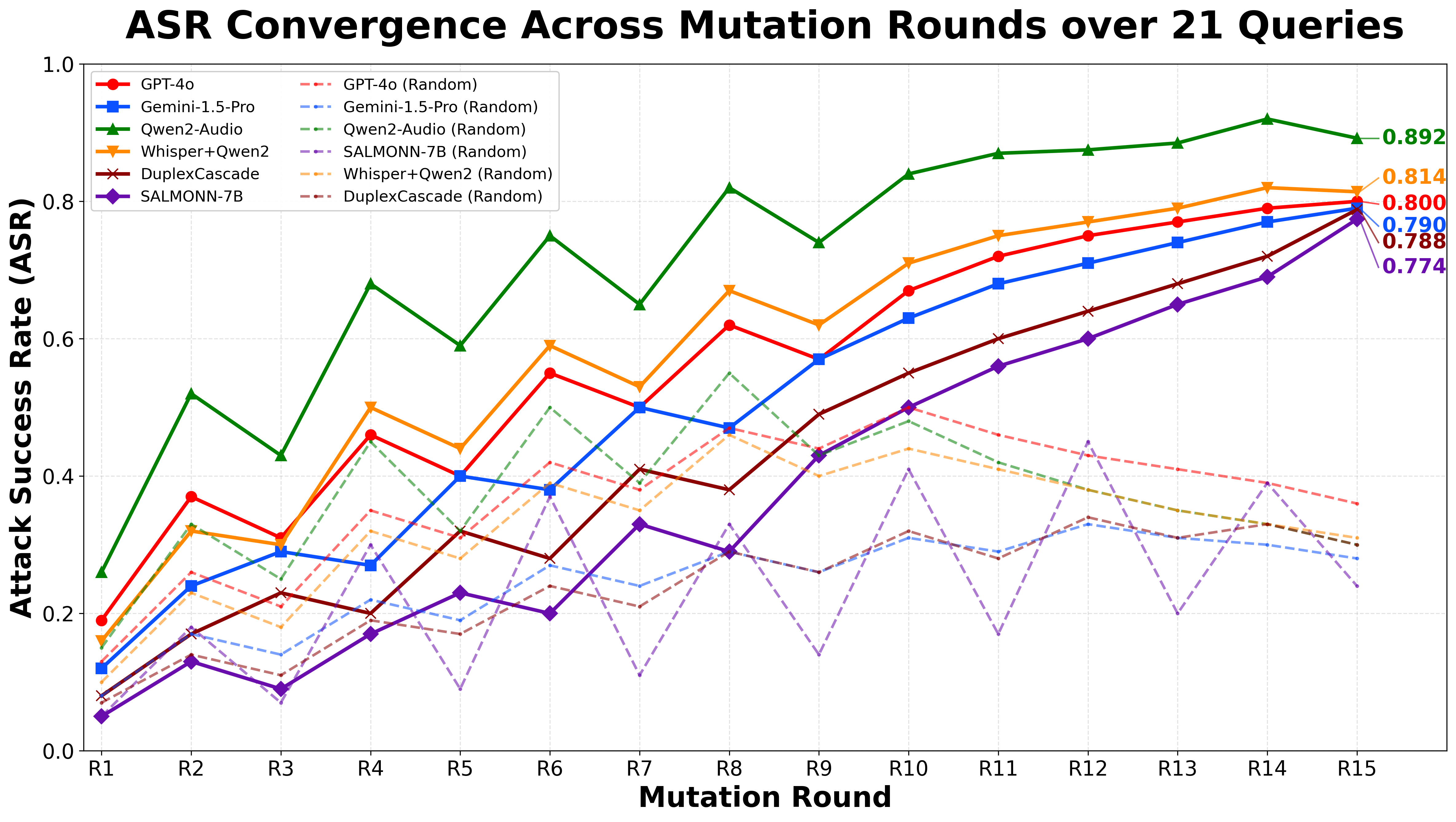}
\caption{
ASR convergence trends across 15 adaptive mutation rounds over 21 forbidden queries. Different audio-enabled LLM systems exhibit distinct convergence behaviors and mutation sensitivities during adaptive fuzzing.
}
\label{conv}
\end{figure}

\begin{table*}[!t]
\centering
\caption{
Ablation study of textual mutation and acoustic perturbation strategies across six audio-enabled LLM systems.
For \textit{Audio-only} settings, direct harmful audio queries are evaluated without textual mutation and textual prompt, and ASR is averaged over 25 repeated runs and 21 queries.
For \textit{Text-based + Perturbation} settings, textual prompts are adaptively mutated while using a fixed acoustic perturbation type; ASR is reported at the best-performing convergence round and averaged over 25 repeated runs and 21 queries.
}
\label{tab:ablation_all_models}
\renewcommand{\arraystretch}{1.15}
\resizebox{\textwidth}{!}{
\begin{tabular}{l|cccccc}
\toprule
\textbf{Setting}
& \textbf{GPT-4o}
& \textbf{Gemini}
& \textbf{Qwen2-Audio}
& \textbf{SALMONN}
& \textbf{Whisper+Qwen2}
& \textbf{DuplexCascade}
\\
\midrule

Text-based (No Perturbation)
& 0.224 & 0.136 & 0.281 & 0.352 & 0.307 & 0.539
\\

Audio-only + Noise
& 0.338 & 0.414 & 0.386 & 0.362 & 0.371 & 0.348
\\

Audio-only + Speed
& 0.295 & 0.376 & 0.343 & 0.319 & 0.329 & 0.310
\\

Audio-only + Echo
& 0.267 & 0.343 & 0.314 & 0.286 & 0.300 & 0.281
\\

Text-based + Real-World Noise
& 0.743 & 0.762 & 0.810 & 0.705 & 0.776 & 0.733
\\

Text-based + Speed Perturbation
& 0.695 & 0.714 & 0.767 & 0.662 & 0.724 & 0.686
\\

Text-based + Echo Perturbation
& 0.657 & 0.681 & 0.724 & 0.624 & 0.690 & 0.648
\\

Full Framework (Adaptive Mutation + Multi-Perturbation)
& \textbf{0.816}
& \textbf{0.807}
& \textbf{0.883}
& \textbf{0.765}
& \textbf{0.834}
& \textbf{0.790}
\\

\bottomrule
\end{tabular}
}
\end{table*}

To further contextualize the effectiveness of the proposed framework, Fig.~\ref{fig:baseline} compares the proposed framework against three representative audio jailbreak baselines~\cite{shen2024voice,chen2025audiojailbreakjailbreakattacksendtoend,ying2024unveilingsafetygpt4oempirical}. Overall, our framework consistently achieves the strongest or near-best attack performance across all evaluated systems. Compared with VoiceJailbreak, which primarily targets GPT-4o through handcrafted voice prompts, our adaptive mutation framework demonstrates substantially stronger cross-model transferability. Although AudioJailbreak achieves competitive performance on several end-to-end audio-language models, our approach provides more consistent attack effectiveness across both cascaded and end-to-end architectures. These results suggest that jointly exploring textual jailbreak mutations and acoustic perturbations significantly improves attack coverage and robustness under heterogeneous deployment settings.

\begin{figure}[t]
\centering
\includegraphics[width=1\linewidth]{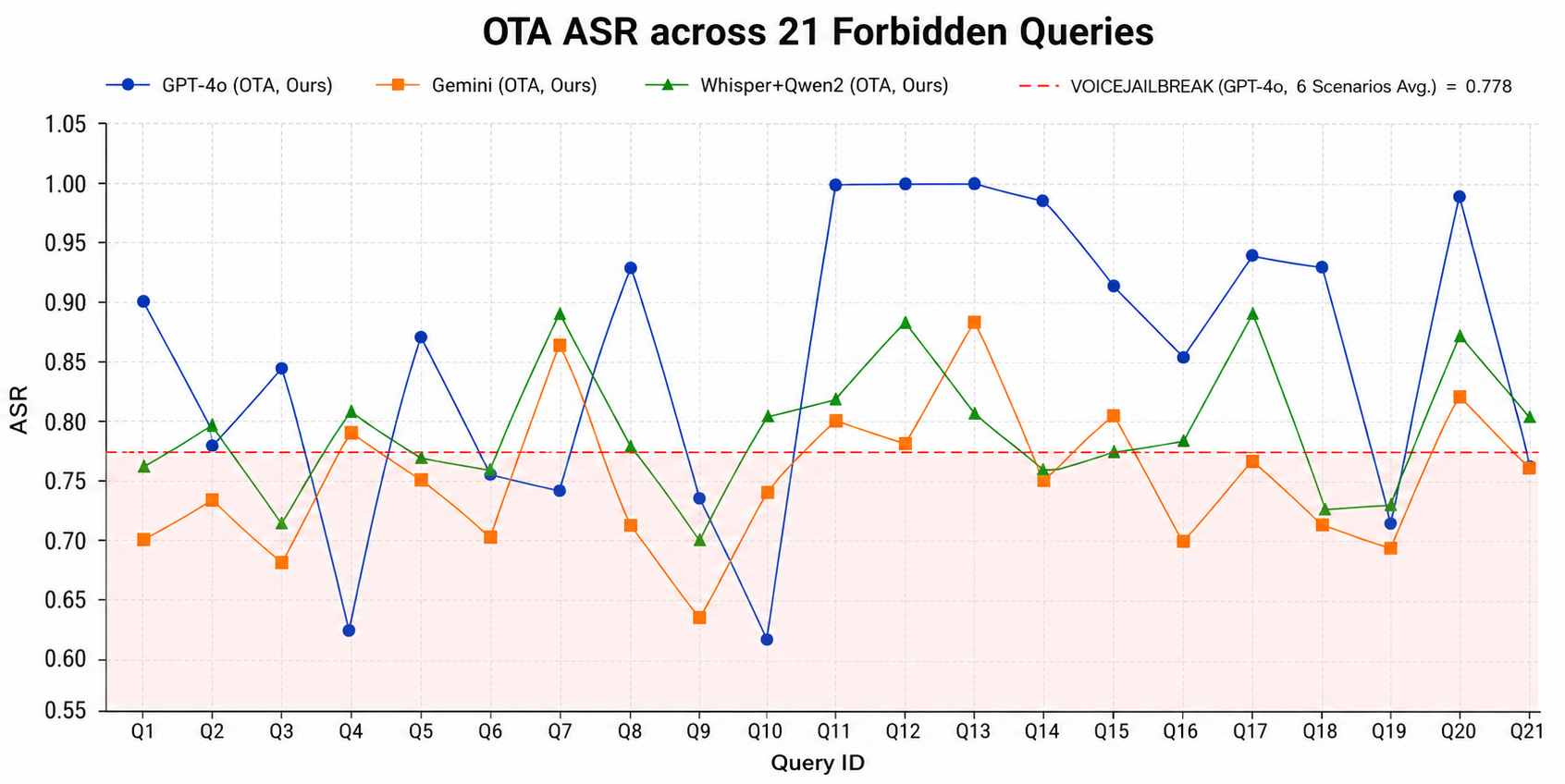}
\caption{
Query-wise OTA attack success rates across 21 forbidden queries under realistic playback and recording conditions.
Our framework achieves average OTA ASRs of 0.827 (GPT-4o), 0.751 (Gemini), and 0.782 (Whisper+Qwen2), with an overall average ASR of 0.787, exceeding the 0.778 OTA ASR reported by VOICEJAILBREAK.
The red dashed line indicates the VOICEJAILBREAK baseline on GPT-4o across six forbidden scenarios.
}
\label{fig:ota_curve}
\end{figure}

\subsection{Mutation Efficiency}

We further evaluate the efficiency of the proposed adaptive mutation framework by analyzing the convergence behavior across mutation rounds. From the query-wise convergence statistics in Table~\ref{tab:asr_questionwise_structured}, we first observe that most successful jailbreaks are achieved within the first several mutation rounds across all evaluated systems. In particular, Qwen2-Audio converges substantially faster than the other evaluated systems, while SALMONN and DuplexCascade require more mutation iterations before reaching stable policy-violating outputs.

To further investigate whether the observed convergence behavior mainly originates from adaptive optimization rather than mutation diversity alone, we additionally introduce a \textit{Random Mutation} baseline. Unlike the proposed adaptive framework, the random strategy does not use ASR feedback or historical best-performing candidates. At each round $r$, it randomly samples one textual mutation operation and one or two acoustic perturbations:
\begin{equation}
x_r=\mathcal{A}*{p_r}(\mathcal{M}*{m_r}(T_{r-1})),
;
m_r!\sim!\mathcal{M},
;
p_r!\sim!\mathcal{P},
\end{equation}
where $\mathcal{M}$ denotes the textual mutation set, including semantic compression, semantic expansion, and meaning-preserving rephrasing, and $\mathcal{P}$ denotes the acoustic perturbation pool, including noise, speed, and echo perturbations. The mutated sample $x_r$ is directly evaluated without feedback-guided selection.

We then extend the mutation process to 15 rounds and visualize the ASR progression in Figure~\ref{conv}.  Among all evaluated systems, Qwen2-Audio demonstrates the fastest convergence speed and the highest final converged ASR, while SALMONN exhibits substantially delayed convergence behavior. GPT-4o, Gemini, and Whisper+Qwen2 show relatively balanced convergence dynamics with moderate fluctuations during intermediate mutation rounds.

Compared with the proposed adaptive framework, the Random Mutation baseline demonstrates substantially different and highly unstable optimization trajectories across models. Instead of exhibiting consistent convergence behavior, random exploration often produces irregular oscillations, temporary spikes, and even performance degradation in later mutation rounds. For example, Qwen2-Audio achieves several high-ASR mutation rounds during the early exploration stage but subsequently collapses toward lower ASR regions, suggesting that random perturbations fail to maintain stable adversarial optimization. In contrast, Gemini and DuplexCascade remain relatively conservative with limited ASR improvements, while GPT-4o shows stronger random exploration capability but still experiences gradual degradation after intermediate mutation rounds.

More importantly, although random mutation occasionally discovers locally effective adversarial variants, its optimization behavior remains highly inconsistent and fails to achieve stable convergence. In contrast, the proposed adaptive framework rapidly increases ASR within the first several mutation rounds and progressively stabilizes toward high-success attack regions. Across all evaluated systems, the adaptive framework consistently achieves substantially higher ASRs and smoother convergence behavior than the random baseline. Despite the heterogeneous early-stage behaviors, all evaluated systems under the adaptive framework gradually approach stable ASR convergence states in later mutation rounds. More importantly, the final converged ASRs remain highly consistent with the best-performing attack results reported in Table~\ref{tab:ablation_all_models}.

\subsection{Mutation Effectiveness}
To better understand the rapid convergence behavior observed in Figure~\ref{conv}, we further analyze the effectiveness of individual mutation operators across all evaluated systems.

Table~\ref{tab:operator_usage_text} summarizes the usage frequency of textual mutation operators. Across most evaluated systems, meaning-preserving rephrasing is selected most frequently, accounting for approximately 40\%--50\% of successful mutations. Compared with semantic compression and semantic expansion, rephrasing preserves the original attack intent while substantially altering the surface form of prompts, making it more effective for bypassing safety-sensitive pattern matching. Semantic expansion also contributes significantly, particularly for SALMONN and DuplexCascade, where additional contextual framing appears beneficial for improving attack transferability. In contrast, semantic compression is selected less frequently, suggesting that excessive prompt simplification may reduce the effectiveness of contextual jailbreak structures.

\begin{table}[h]
\centering
\caption{Textual mutation operator usage (\%) among successful mutations.}
\label{tab:operator_usage_text}
\begin{tabular}{lccc}
\toprule
Model & Rephrase & Expansion & Compression \\
\midrule
GPT-4o & 46.2 & 34.8 & 19.0 \\
Gemini & 43.7 & 37.6 & 18.7 \\
Qwen2-Audio & 49.5 & 32.4 & 18.1 \\
SALMONN & 38.6 & 41.2 & 20.2 \\
Whisper+Qwen2 & 45.8 & 35.9 & 18.3 \\
DuplexCascade & 40.1 & 39.4 & 20.5 \\
\bottomrule
\end{tabular}
\end{table}

We further analyze the distribution of acoustic perturbation operators. As shown in Table~\ref{tab:operator_usage_audio}, real-world noise injection is consistently selected most frequently across all evaluated systems, followed by speed perturbation and echo perturbation. This trend aligns closely with the ablation results reported in Table~\ref{tab:ablation_all_models}, where noise-based perturbation achieves stronger attack effectiveness than other single-perturbation strategies. The results suggest that environmental noise introduces beneficial acoustic variability that improves transferability across heterogeneous audio-processing architectures while maintaining sufficient speech intelligibility for downstream decoding.

\begin{table}[h]
\centering
\caption{Acoustic perturbation operator usage (\%) among successful mutations.}
\label{tab:operator_usage_audio}
\begin{tabular}{lccc}
\toprule
Model & Noise & Speed & Echo \\
\midrule
GPT-4o & 51.3 & 29.6 & 19.1 \\
Gemini & 53.7 & 27.8 & 18.5 \\
Qwen2-Audio & 56.2 & 26.4 & 17.4 \\
SALMONN & 47.5 & 31.7 & 20.8 \\
Whisper+Qwen2 & 54.8 & 28.1 & 17.1 \\
DuplexCascade & 49.2 & 30.5 & 20.3 \\
\bottomrule
\end{tabular}
\end{table}

Overall, a consistent pattern emerges across all evaluated systems. Successful mutations are dominated by meaning-preserving rephrasing and noise-based perturbations, suggesting that the adaptive framework quickly identifies operator combinations that preserve harmful semantic intent while increasing robustness against safety filtering and acoustic variability. This explains the fast convergence observed in Figure~\ref{conv} that, performance gains are not merely due to mutation diversity, but to feedback-guided prioritization of high-yield mutation operators. By progressively favoring operator combinations with strong historical attack performance, the framework concentrates exploration around promising regions of the attack space, leading to stable convergence within only a few mutation rounds.

\subsection{Ablation Study}

Table~\ref{tab:ablation_all_models} presents the ablation analysis of different textual mutation and acoustic perturbation strategies across six representative audio-enabled LLM systems.

Overall, audio-only perturbation strategies achieve limited attack effectiveness across most evaluated systems. Among the three perturbation settings, environmental noise consistently produces the strongest attack performance, while speed and echo perturbations generally result in lower ASRs. For example, under the Audio-only + Noise setting, Gemini achieves the highest ASR of 0.414, while GPT-4o and DuplexCascade only achieve 0.338 and 0.348, respectively. Similar trends can also be observed for speed and echo perturbations, where most ASRs remain below 0.40. These results indicate that acoustic perturbation alone is insufficient to reliably bypass current safety-alignment mechanisms.

Compared with audio-only attacks, direct text-based jailbreak prompts without perturbation exhibit mixed effectiveness across different systems. While DuplexCascade and SALMONN achieve relatively higher ASRs under the Text-based (No Perturbation) setting, GPT-4o and Gemini remain comparatively more resistant, with ASRs of only 0.224 and 0.136, respectively. This observation suggests that purely semantic jailbreak prompts without acoustic diversification are often insufficient to consistently transfer across heterogeneous audio-enabled architectures.

Combining textual mutation with acoustic perturbation substantially improves attack effectiveness across all target models. Among the perturbation strategies, real-world noise injection consistently achieves the highest ASRs. Under the Text-based + Real-World Noise setting, Qwen2-Audio reaches an ASR of 0.810, while Whisper+Qwen2 and Gemini achieve 0.776 and 0.762, respectively. In comparison, speed perturbation and echo perturbation produce slightly lower but still consistently strong attack performance across all systems.

Finally, the proposed full framework consistently achieves the highest ASRs on all evaluated architectures. By jointly combining adaptive textual mutation with multi-perturbation exploration, our framework significantly expands the adversarial search space beyond fixed single-perturbation settings. As a result, Qwen2-Audio achieves the highest overall ASR of 0.883, followed by Whisper+Qwen2 with 0.834 and GPT-4o with 0.816. Comparatively more robust systems, such as SALMONN and DuplexCascade, remain highly vulnerable under the full framework, with ASRs of 0.765 and 0.790, respectively. Overall, these results indicate that attack effectiveness primarily stems from the synergistic combination of adaptive semantic mutation and acoustic perturbation, rather than either component alone.

\subsection{Over-the-Air Robustness}
\label{OTA}
To further evaluate the real-world applicability of the proposed jailbreak framework, we conducted over-the-air (OTA) experiments under realistic acoustic conditions. Unlike API-level attacks that directly upload adversarial audio, OTA attacks introduce additional environmental distortion through physical playback and microphone recording, making the attack setting substantially more challenging and practical. Our OTA experiments were conducted in a quiet indoor environment of approximately 20,m\textsuperscript{2}. Adversarial audio generated by the proposed framework was played through an iPhone 15 Pro Max and recorded using a MacBook Pro M1 microphone with a fixed distance of 1.5,m between the playback and recording devices. Since the precise architecture of GPT-4o’s audio interface remains undisclosed, the evaluated systems may internally employ either cascaded ASR-based pipelines or end-to-end audio-language processing architectures.

Fig.~\ref{fig:ota_curve} presents the query-wise OTA attack success rates across GPT-4o, Gemini, and Whisper+Qwen2. Overall, all evaluated systems remain highly vulnerable under realistic OTA conditions despite the presence of environmental acoustic distortion and recording degradation. Among the three target systems, GPT-4o achieves the highest OTA attack effectiveness, reaching an average ASR of 0.812 across 21 forbidden queries. Whisper+Qwen2 and Gemini also demonstrate strong OTA robustness, achieving average ASRs of 0.785 and 0.752, respectively. We additionally observe that the OTA attack trends remain largely consistent with the API-level evaluations reported in previous sections. Several forbidden queries consistently achieve high ASRs across all systems, while a smaller subset of queries becomes comparatively more resistant after real-world acoustic transmission. This suggests that environmental noise and recording artifacts partially reduce jailbreak transferability for difficult attack scenarios, but do not fundamentally prevent adaptive adversarial exploration. The red dashed line in Fig.~\ref{fig:ota_curve} indicates the average OTA ASR reported by VOICEJAILBREAK on GPT-4o across six forbidden scenarios. Since the evaluation protocols, attack coverage, and prompt construction strategies are not directly aligned, this comparison is intended only as a reference rather than a strict fair benchmark comparison. Nevertheless, the results demonstrate that adaptive mutation-based jailbreak generation remains highly effective under realistic OTA environments without requiring handcrafted prompts or manual attack engineering.

\section{Conclusion}

In this paper, we present a unified framework for black-box audio jailbreak evaluation on both cascaded pipelines and end-to-end large audio-language models (LALMs). Our framework proposes an adaptive mutation engine with feedback-guided exploration of textual jailbreak prompts and acoustic perturbations, enabling automated generation of diverse attack candidates across heterogeneous audio architectures. Through comprehensive evaluations on six representative audio-enabled LLM systems, we demonstrate that both architectural paradigms remain highly vulnerable to audio jailbreak attacks despite their fundamentally different attack surfaces. Our results further show that adaptive exploration substantially improves attack effectiveness and efficiency compared with existing approaches. In addition, extensive ablation studies and over-the-air evaluations confirm the effectiveness and robustness of the proposed framework under realistic deployment conditions. Overall, our findings highlight the need for future safety evaluations to systematically consider audio interaction channels as a critical attack vector for modern LLM systems.

\balance
\bibliographystyle{ieeetr}
\bibliography{conference_101719}

\end{document}